\documentstyle[epsf]{npbproc}

\font\twelvemsym=msym10 scaled 1200
\font\tenmsym=msym10
    \newfam\msymfam
    \def\msym{\fam\msymfam\twelvemsym}
    \textfont\msymfam=\twelvemsym
    \scriptfont\msymfam=\tenmsym
    \scriptscriptfont\msymfam=\scriptfont\msymfam
\font\tenmib=cmmib10
\font\ninemib=cmmib10 scaled 900
    \newfam\mibfam
    \def\mib{\fam\mibfam\tenmib}
    \textfont\mibfam=\tenmib
    \scriptfont\mibfam=\ninemib
    \scriptscriptfont\mibfam=\scriptfont\mibfam

\newcommand{\be}{\begin{equation}}
\newcommand{\ee}{\end{equation}}
\renewcommand{\l}{\mathop{\cal L}}
\newcommand{\ul}[1]{\mathop{\underline{#1}}}
\newcommand{\G}{\mathop{\cal G}}
\renewcommand{\tiny}[1]{\mathop{\scriptscriptstyle#1}}
\newcommand{\NN}{{\msym N}}
\newcommand{\CC}{{\msym C}}
\renewcommand{\vec}[1]{\mbox{\mib #1}}

% ...

\begin{document}
\title{PATH INTEGRALS AND VORONIN'S THEOREM ON THE
UNIVERSALITY OF THE RIEMANN ZETA FUNCTION} 
\author{Khalil M. Bitar}
\address{Supercomputer Computations Research Institute,
Florida State University,
Tallahassee, FL, USA 32306-3006}

\date{}

\runtitle{Path Integrals and Veronin's Theorems} 
\runauthor{K.M. Bitar} 
\volume{XXX}  
\firstpage{1} 
\lastpage{3}  

\begin{abstract}
We explore a new approach to the path integral for a latticized quantum
theory. 
\end{abstract}
\maketitle
This talk is based on work with N.~Khuri and
H.~Ren.\cite{bitar1,bitar2} 

 The main new tools we use are theorems by Voronin on the
universality properties of the Riemann Zeta function, $\zeta(s)$, in the
critical strip, ${1\over 2}< Re\; s <1$.

Given any real continuous function,
$\phi(x),\;0\leq x \leq \L$, we can choose {\bf a} mapping, $s(x)$, which
maps the line, $0 \leq x \leq \L$, onto a line $s(x)$ that lies in the
strip ${1\over 2} < Re\; s <1$, then given any $\Delta > 0$, we have an
infinite set of integers, $\l$, such that for all $n\epsilon\l$, 
\be
\left | \phi(x) - \gamma\,(s(x) +  in\,\Delta)\right | < \varepsilon,\;
\;\; 0\leq x \leq \L 
\ee
here $s(x)$ is fixed and could be taken to be linear, $s(x)={1\over 2} +
{x\over 2\L}$; and $\epsilon$ is arbitrarily small, $\gamma=\ln|\zeta|$.

Thus for all integers $n$ the functions $\left \{ \gamma\,
(s(x_j)\right.$ $\left. + in\,\Delta)\right \}$ will come arbitrarily close
to ``all paths'' $\left \{\phi(x_j)\right \}, \;\;0 \leq x, \leq \L,\;\;
j=1,\,2,\,\ldots,\nu$. 

The path integral may be written as a discrete sum over
paths labeled by an integer, $n$, and given by
$\{\gamma_i (s(x_j) + in\,\Delta)\}\;\;,j=1,\,\ldots,\nu$.

One then has to introduce a measure, or density function, $\rho_{\nu}(n)$,
which essentially counts the number of different integers ``$n_{\ell}$''
which give paths equivalent to $n$. 

\section{Path Integrals}
In quantum mechanics or quantum field theory one has to evaluate integrals 
of the following from:
\begin{eqnarray}
   \left<  {\rm P}(\phi) \right>&=&
         {1\over \Omega} \int
          \prod^{\nu}_{j=1} d \phi (j) \; e^{-{\rm S}(\phi(j))}\nonumber\\
               &&\times {\rm P} \left(
                \phi(\ell_1)\ldots\phi(\ell_m)
                \right) \,,  \\
\noalign{\hbox{with}}\nonumber\\
\hfil\Omega&=&\int \prod^{\nu}_{j=1} d \phi(j)\;
          e^{-{\rm S} (\phi(j))}
\end{eqnarray}

Here $\nu$ is the number of lattice points, $\phi(j)$ is the field at
the {\em j}'th lattice point, ${\rm S}(\phi)$ is the Euclidean action, and
${\rm P}(\phi (\ell_1)\ldots\phi(\ell_m))$ is a polynomial in the fields.

Such an integral can be easily well approximated by a summation.

{\em We shall assume 
that $\rho_{\nu}$ exits and is a smooth function of $\gamma$.}

\section{Properties of the Measure $\bf{\rho(n)}$}
We concentrate here on the central question of this talk, namely 
the properties of the ``density'' or ``measure'', $\rho_{\nu}(n)$. Indeed we 
shall derive an explicit integral representation for it.

The first thing to do is to actually plot the distribution of values of 
zeta functions. We 
consider the set of zeta functions:  $\zeta (\sigma + in\,\Delta + 
i{\rm T}_0); \;\;n=1,2,\ldots,{\rm N};$ ${1\over 2} < \sigma < 1$; any real 
$\Delta > 0$ (actually $\Delta > 10$, see below), and a fixed 
${\rm T}_0 > O(10^6)$  and compute their values.

This computation strongly indicates that there exists a probability 
distribution, ${\rm W}_\sigma(\gamma)$ independent of $\Delta$ and of ${\rm
T}_0$ (for large ${\rm T}_0$), which tells us that if we randomly choose a
$t > {\rm T}_0$ and calculate $\gamma(\sigma + it)$, the probability that $u
<\gamma < u + du$, is given by ${\rm W}_\sigma (u)du$.

  We define our $n$'th ``path" by the
configuration
\begin{eqnarray}
\phi^{(n)}(x_j)&\equiv&\log\left|\zeta(\sigma + ijh +in\Delta)\right|
= \gamma_{\sigma}(j,n)\nonumber\\
&&~~j=1,...,\nu.
\end{eqnarray}
Here to separate our paths we take $\Delta\gg h$, and $\Delta > h\nu$.

It now follows from the standard definition of the path integral 
that one 
can write
\be
\Omega(\nu)=\sum_{n=N_0}^N{ e^{-S(n;\nu)}\over {\rho_\nu(n)} } 
+ {\cal O}(N^{- {1\over \nu}}).
\ee

We must take $N$ large, and $N\gg     N_0$.  By summing over $n$ we sum
over all ``paths", but the density $\rho_\nu(n)$ insures that we have the 
correct Jacobian for quantum mechanics. 

The main question here is to find the expression for,
$\rho_\nu(n)$.

We first discuss the case $\nu=1$.  Here $\rho_1(n)$ can be computed
once we know the asymptotic probability for having $|\zeta(\sigma+it)|$, for
a randomly chosen $t\gg  1$, be such that $r_1 < |\zeta(\sigma+it)| < r_2$.
We call this probability density, ${\cal P}_\sigma(r)$, and write
\begin{eqnarray}
Prob.\left(r_1 <\left|\zeta(\sigma + it)\right|<r_2\right)&\equiv&
\int_{r_1}^{r_2}{\cal P}_\sigma (r)dr,\nonumber\\ 
&&{1\over 2}<\sigma<1. 
\end{eqnarray}

The moments of ${\cal P}_\sigma(r)$ are known for $0\leq  Re~k\leq  2$,

\begin{eqnarray}
\int_0^\infty{\cal P}_\sigma (r) r^{2k} dr &\equiv& \lim_{T\rightarrow
\infty}{1\over T}\int_1^\infty \left|\zeta (\sigma + it)\right|^{2k}dt
   \nonumber\\
&&=F_k(2\sigma)
\end{eqnarray}
and
\be
F_k(2\sigma)=\prod_p~~_2F_1(k,k,1,{1\over {p^{2\sigma}}})
\ee
where $_2F_1(a,b;c;z)$ is the standard hypergeometric function, and $ \Pi_p$
is a product over all primes.   We can obtain an explicit expression
for ${\cal P} _\sigma(r)$ by taking the inverse Mellin transform and get
\be
{\cal P}_\sigma (r)=r^{-c}\int _{-\infty}^{+\infty}d\lambda ~(r)^{-i\lambda} 
F_{ {i\lambda\over2} + {(c-1)\over2}}(2\sigma) 
\ee
where we can take any real c such that $1< c < 5$.  The 
integral above is absolutely convergent.  It is now
easy to obtain the distribution function of the values of $\log|\zeta |$,
which we call $W_\sigma(\gamma)$,
\be
Prob.\bigl (\gamma_1 <|\zeta(\sigma + it)|<\gamma_2\bigr )\equiv 
\int_{\gamma _1}^{\gamma _2}W_\sigma (\gamma)d\gamma
\ee
for a randomly chosen $t\gg   1$.  It is clear that 
$W_\sigma(\gamma)\equiv {\cal P}_\sigma (e^\gamma).e^\gamma$ , and hence
\begin{eqnarray}
W_\sigma(\gamma)&=&e^{-(c-1)\gamma} 
\int_{-\infty}^{+\infty}\nonumber\\ 
&&\qquad d\lambda ~e^{-i\lambda\gamma} 
F_{ {i\lambda\over2} + {(c-1)\over2}}(2\sigma)
\end{eqnarray}
with $1 < c  <5$.  The probability density $W_\sigma(\gamma)$ is an
asymptotic distribution in the sense that if we take an interval $T_0 < t <
T, T\gg   T_0$, and compute a large ensemble of $\log|\zeta(\sigma +
it_j)|$ values, then the resulting histogram for the distribution of values
for this ensemble will approach $W_\sigma(\gamma)$ as $T\rightarrow\infty$,
and $T_0$ is kept fixed.  However, it is fortunate that even for an
interval  $T_0 = {\cal O}(10^6)$ and  $T={\cal O}(10^9)$ the computed
histogram and the exact result for $W_\sigma (\gamma)$ are quite close to
each other. This fact is shown in the figure where the region $T_0 <t < 1 0^9$
which is easily accessible to computer is probed. 
\begin{figure}
\vspace{2.15in}
\includegraphics{fig1.eps}
\caption{Distribution of the values of $\gamma = ln|\zeta|$ at $\sigma=0.75$.  
The crosses denote points computed from the exact formula, and the line is
the histogram from computing a large sample.} 
\end{figure}
The moments of $W_\sigma (\gamma)$ can be given exactly.\cite{bitar2}

\section{\bf The Measure for Multiple Variables}

For $\nu =1$, the density $\rho_1(n)$ is simply given by
 $\rho_1(n)= W_\sigma (\gamma (1;n))$, with $\gamma (1;n)$ defined
in eqn.~(4).  For $\nu > 1$, and $h >1$, the values of $\gamma_\sigma
(j;n)$ and $\gamma_\sigma (j+l ;n)$ are uncorrelated. Thus the probability
density is the product of the independent probabilities for each component.
 This leads to a factorization of $\rho_\nu (n)$ and we get
\be
\rho_\nu(n)=\prod_{j=1}^\nu W_\sigma(\gamma_\sigma(j;n))
\ee

We now have, for example, the following explicit formulae for the partition
function in Euclidean Quantum Mechanics:
\begin{eqnarray}
\Omega(\nu)&=&\sum_{n=1}^N e^{-S(n;\nu)}\left [\prod_{j=1}^\nu
 W_\sigma \left(\gamma_\sigma (j;n)\right)\right]^{-1}\nonumber\\
&&+{\cal O}(N^{- {1\over\nu}}) 
\end{eqnarray}
For results of numerical tests, we refer to References [1] and [2].

\section{\bf Conclusions and Remarks}
We end this paper with a miscellaneous set of remarks and give
projections on further work.  
\begin{itemize}
\item[(i)]  {The calculations we have carried out
demonstrate the validity of the expressions for the partition function in
Euclidean quantum mechanics.  This includes the validity of the factorization
conjecture for the density $\rho_\nu(n)$. }

\item[(ii)]  The applications of our
method could either be for actual numerical calculations or for obtaining
formal results.   It may also be used for a complex actions.

\item[(iii)]  The continuous form of Voronin's theorem, 
leads us to contemplate a far reaching conjecture.  This
concerns taking the limit $a\rightarrow 0$, where $a$ is the lattice
spacing.  If in this limit a measure, $\rho_\infty (n)$, exists, then
essentially any quantum mechanical problem can be reduced to quadratures. 
\end{itemize}

The important thing to remember is that $\rho_\infty (n)$ depends only on
the properties of the Riemann zeta function. All
the physics enters through $S (n)$.  Clearly, the existence of $\rho_\infty
(n)$ and an explicit formula for it would be a remarkable achievement. 


\end{document}